\documentclass{elsart}

\usepackage{graphicx}
\usepackage{amssymb}

\begin{document}

\begin{frontmatter}

\title{Radiation shielding for
  underground low-background experiments}

\author{D.Y. Stewart\corauthref{cor}},
\corauth[cor]{Corrresponding author.}
\ead{d.y.stewart@warwick.ac.uk}
\author{P.F. Harrison}, 
\author{B. Morgan}, 
\author{Y. Ramachers}

\address{University of Warwick, Coventry, CV4 7AL, UK}

\begin{abstract}
The design task of creating an efficient radiation shield for the new
COBRA double-beta decay experiment led to a
comprehensive study of commercially available shielding materials. The
aim was to find the most efficient combination of materials under the
constraints of an extreme low-background experiment operating in a
typical underground laboratory. All existing shield configurations for
this type of experiment have been found to perform sub-optimally in
comparison to the class of multilayered configurations proposed in
this study. The method used here to create a
specific shield configuration should yield a close to optimal
result when applied to any experiment utilising a radiation shield. In
particular, the survey of single material response to a given
radiation source turns out to give a guideline for the
construction of efficient multilayer shields. 
\end{abstract}

\begin{keyword}
double-beta decay \sep radiation shielding \sep Monte-Carlo \sep dark
matter \sep low background
\PACS 23.40.-s \sep 24.10.Lx \sep 28.20.Gd 
\end{keyword}
\end{frontmatter}

\section{Introduction}\label{intro}

The study of radiation shielding has a large interdisciplinary
scope, ranging from high-energy physics accelerator infrastructure to
medical physics and engineering science \cite{conf}. For generic low-background
experiments utilising radiation detectors of any kind, a radiation
shield is a major part of the set-up \cite{heusser}. Rare event
searches, as in particle dark matter and double-beta decay search
experiments, rely heavily upon an extremely low level of background
contributions from environmental radiation sources. So too do experiments
measuring tiny radioactive contaminations of samples, e.g. in
archaeology and tracer searches in oceanography. 

\par
In the framework of the authors' main research project, the COBRA double-beta
decay experiment \cite{cobra}, the focus was on the design of a new
radiation shield as one of the major constructional parts for the
success of the current R\&D efforts into COBRA. A survey of existing
shields for similar experiments soon showed that hardly any details are
published, although they always constitute a major part of the physics
programme. Therefore, it was decided that the issue should be investigated independently to see whether it would be viable to improve on existing
designs, not only for the specific case of the COBRA experiment but in
general. 

\par
The following strategy has been implemented for this study: 
\begin{itemize}
\item Constrain the database of viable shielding materials to
  commercially available and mechanically convenient samples. A typical radiation shield weighs in excess of several
  hundred kilograms, hence cost and construction are important
  factors. 
\item Examine each single material for its radiation interaction
  properties for all relevant radiation sources. In our case, these
  consist of $\gamma{}$-radiation and neutrons. It should be pointed
  out that this study concentrates on neutron radiation and that 
  $\gamma{}$-radiation mainly becomes a concern when originating from
  inelastic neutron reactions, rather than in isolation. In underground
  laboratories, the hadronic component of cosmic rays is absent and
  for the deeper laboratories, the muonic component is suppressed
  compared to environmental radioactivity
  \cite{heusser}. 
\item Maximise the attenuation for a specific radiation source
  by combinations of suitable materials, building a multilayer
  shield. The
  motivation for this process originates from general characteristics
  of shielding, i.e. (a) scatter particles to
  lower energies and (b) absorb particles, where (b) is always a
  stronger effect at lower energies. However, this rather simple
  few-parameter problem is complicated by the fact that
  particle number is generally not conserved for radiation interaction
  in matter. Build-up factors \cite{harima} are important and render the whole problem tractable only
  numerically, i.e. using Monte-Carlo simulations. 
\item Maximise the attenuation for any combination of radiation
  sources. It is understood that this might not always be necessary
  for every experiment, especially those that are
  insensitive to certain sources due to strong discrimination
  capabilities, as in dark matter experiments (see for instance
  \cite{dm}).
\end{itemize}
This text then concludes with a comparison of our proposed multilayer
shields with `standard' shields, employed in existing underground
experiments.

\section{Single Materials Survey}\label{sec1}

The motivation to focus this study on neutron sources is
twofold. First, studies and data on the subject of $\gamma{}$-radiation on
various shielding materials exist (see \cite{heusser}, \cite{harima}, 
\cite{shizuma} and references therein), even with the scope
of examining multilayer shields \cite{bisselle}.
 Second, the COBRA
experiment, utilising CdZnTe semiconductor detectors will be
particularly sensitive to thermal neutron capture reactions on
$^{113}$Cd, producing $\gamma{}$-radiation in the relevant signal
energy region (above two MeV). 
\par

\begin{table}
\caption{Shielding materials and parameters used in the
  simulations. Note that mixtures are typically given as weight
  percent, whereas an atom number ratio is more useful for
  specifying materials for simulations. Natural isotope composition is
  always assumed.\label{tab1}}
\vspace*{6pt}
\begin{center}
\begin{tabular}{lll}
\hline
Material & Composition [atom number ratio]& Density [g/cm$^{3}$]\\\hline
Lead & nat.& 11.34\\
Iron & nat.& 8.96\\
Copper & nat.& 7.87\\
PE & CH$_{2}$[1:2]& 0.92\\
PE-Bi$^{1}$ & CH$_{2}$,Bi [1:2:0.14]& 3.0\\
PE-B30\%$^{1}$ & CH$_{2}$,B [1:2:0.6739]& 1.12\\
PE-B5\%$^{1}$ & CH$_{2}$,B [1:2:0.0788]& 0.95\\
PE-Li$^{1}$ & CH$_{2}$,Li [1:2:0.1231]& 1.06\\
Premadex$^{2}$ & HCOLi [1:0.35:0.2217:0.0162]& 1.0\\
Water & H$_{2}$O [2:1]& 1.0\\\hline
\multicolumn{3}{l}{$^{1}$ {\small Thermo-Electron Corporation;} $^{2}$
  {\small Wardray Premise Group}}\\
\end{tabular}
\end{center}
\end{table}

\par
Several recent studies for underground experiments quantified design
criteria for shielding; in particular neutron shielding was targeted
due to interest in new dark matter search experiments
\cite{drift1,drift2,hesti}. However, none of those experiments aimed at a
complete description i.e. including $(n,\gamma{})$-reactions and shielding
effects on the combined neutron and photon population. 

\par 
For this study, commercially available shielding
materials were selected, as summarised in Table ~\ref{tab1}. Most materials tabulated here are specialised neutron shielding materials (the hydrocarbon class) but included are a
few pure metals. As a representation of typical neutron fluxes
as function of energy in a deep underground laboratory, the
measured neutron flux 
from Gran Sasso was adopted \cite{belli} (see also the updated discussion about
Gran Sasso neutron fluxes \cite{hesti2}), see Table ~\ref{gransasso}.  

\par
\begin{table}[htb]
\begin{center}
\caption{\small Measured neutron flux as function of the indicated
energy bins, from \protect{\cite{belli}}. \label{gransasso}}.
\vspace*{3mm}
\begin{tabular}{ll}
\hline
Energy bin [MeV] & n-flux\\
 & ($10^{-6}$cm$^{-2}$s$^{-1}$)\\\hline
$<50\times{}10^{-9}$ & 1.07$\pm$0.05\\
$50\times{}10^{-9} - 10^{-3}$ &1.99$\pm$0.05 \\
$10^{-3} - 2.5$ & 0.53$\pm$0.08 \\
2.5 - 5 &0.18$\pm$0.04 \\
5 - 10 &0.04$\pm$0.01 \\
10 - 15 &(0.7$\pm$0.2)$10^{-3}$ \\
15 - 25 &(0.1$\pm$0.3)$10^{-6}$ \\\hline
\end{tabular}
\end{center}
\end{table}
\par

\par
The main tool for this research is computational dosimetry using
Monte-Carlo codes. Results are predominantly based on the MCNPX code
package \cite{mcnpx} and confirmation was sought from semi-independent
simulation packages based on the GEANT4 framework \cite{geant}, (see \cite{Lemrani} for more information). Application code was written using the GEANT4 framework. 
Both codes, MCNPX and GEANT4,  essentially use identical cross section
libraries for the energetically relevent
neutron interactions in matter (below 20 MeV neutron energies); hence, they are referred to as semi-independent even though all other algorithms are different. 
\par
\begin{figure}
\includegraphics[width=14cm]{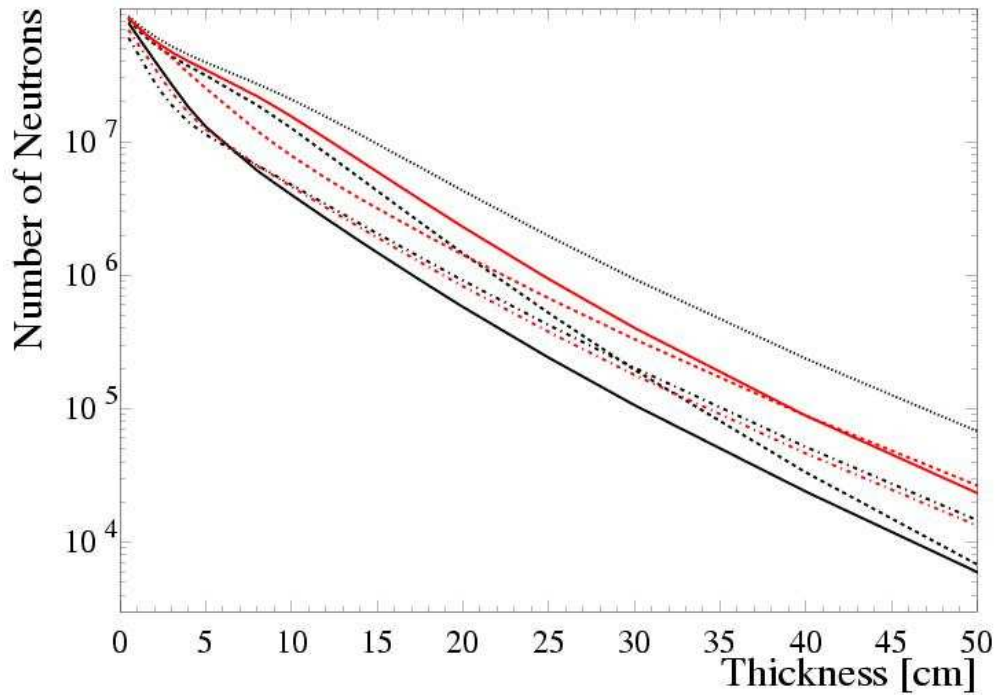}
\caption{Neutron shielding effects shown as a function of absorber
  thickness for various bulk materials. From top to bottom (at 15 cm
  thickness): Water (dotted), pure polyethylene (PE, solid grey),
  PE with bismuth (dashed), Premadex (grey dashed), PE with 30\%
  boron content and 5\% boron content (dash-dotted, B30\% in black,
  B5\% grey) and PE with lithium (solid).\label{fig1}}
\end{figure}

\par
An indication of the shielding effect of various neutron
absorber materials can be seen in Fig.~\ref{fig1}. The boronated
polyethylene (PE) shields perform best in terms of neutron number reduction
for thin shields ($<5$cm). After 5cm thickness, the PE+Lithium shield 
excels. The hydrocarbons containing admixtures of neutron capture
materials clearly dominate pure PE in terms of attenuation but it
should be noted that this effect is due to thermal neutron numbers
being significantly reduced and not due to an overall effect on the
full neutron population. The 
cheapest shielding material, water, unfortunately also turns out to be
the worst in performance. For the mixture hydrocarbon materials the
cost factor is a non-negligible constraint, particularly for larger
thicknesses. 

\par
However, the observable total neutron number alone is often not the most
important for experiments interested in energy
measurements. It appears that studies focussing on total
neutron number have served, thus far, to define design criteria for
shields. Here it is argued that the energy dependence of the penetrating flux 
is at least as important as number counts; in addition, build-up
factors from neutron inelastic reactions either in terms of neutron
multiplication or $\gamma{}$-production are other important factors to
consider. 

\begin{figure}
\includegraphics[width=14cm]{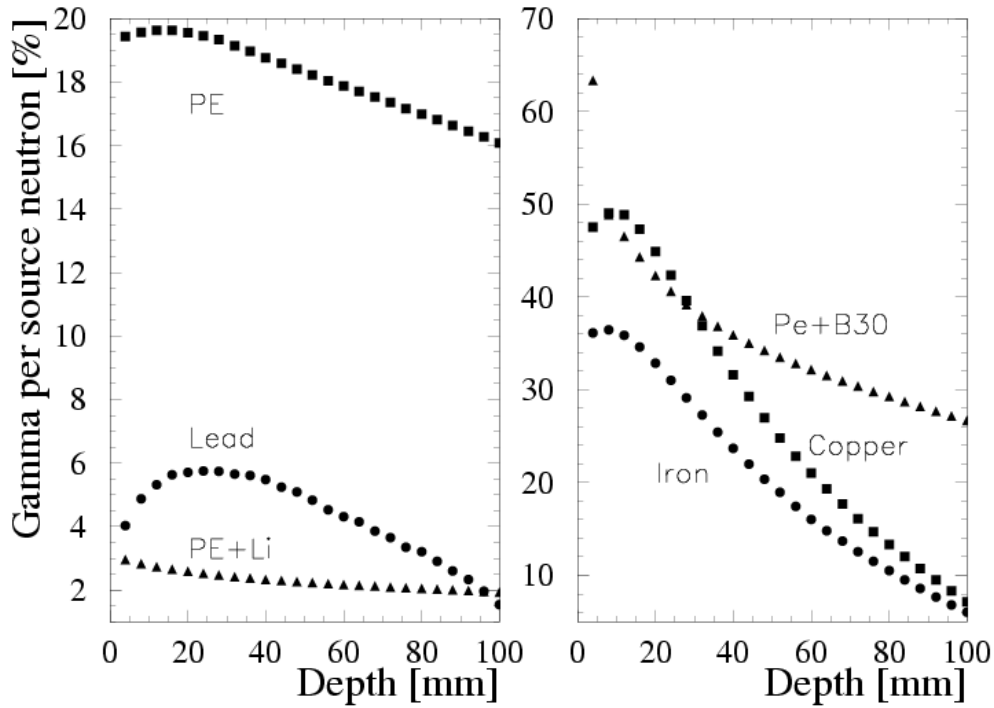}
\caption{$\gamma{}$-ray population per source neutron as a function of material
thickness. A total depth of 10cm has been divided into 4mm
slices. Shown is the gamma population, i.e. the number of photons
present in each slice including actually produced photons and
transported photons from 
other locations. As source, a pure neutron beam with Gran Sasso energy spectrum
has been used. As bulk materials, pure PE, lead, PE+Lithium (all
on the left panel), copper, iron and PE+30\%Boron (right panel) have
been chosen.\label{fig2}}
\end{figure}

\par
Of particular interest for this study is the $(n,\gamma)$-reaction on
metals and hydrocarbons and also the bulk material
self-absorption effect. To investigate, bulk materials of copper, iron, lead, pure
polyethylene and the mixtures PE+Li and PE+Boron(30\% enriched) have
been defined to have a 10cm thickness and have been sliced into 4mm
pieces. Photon population number in each 4mm slice was registered even
though a pure neutron beam source (with Gran Sasso spectrum) was
started. The results can be inspected in Fig.~\ref{fig2}. 
\par

\par
The $\gamma{}$-population per source neutron as a function of material
thickness confirms the usual assertion that lead is a superior
shielding material not only for $\gamma{}$-radiation but also due to
its small reaction cross sections with neutrons. The overall
$\gamma{}$-production from lead is far smaller than from all other
metals in this study, reaching roughly 6\% at
maximum. Pure PE can go up to 20\% but standard shielding metals like
copper and iron reach well above plastics, almost to 50\% in the case
of copper. The hydrocarbon-based materials display a remarkable feature that
should be bourne in mind when applying them to neutron shielding. Despite comparable performance with respect to neutron moderation and capture, there
nevertheless is a huge difference in terms of photon
production. The Boron admixture produces a vast amount of photons,
particularly in the first slice, i.e. in the first 4mm. The lithium
mixture shows very little photon production, in fact even fewer than lead, and
singles itself out as a potential ideal capture material close to
detectors.

\par
Self-absorption of $\gamma{}$-rays can also be seen from
this figure. PE and its mixtures hardly re-absorb their own
$\gamma{}$-radiation,
whereas the metals are more efficient. 
Iron produces a maximum number of photons around 36 - 37\% compared to
just 6\% for lead. After 10cm thickness, lead has reduced its
value of gammas per source neutron below 2\% but copper needs more than 10cm to compete. These
results for inelastic neutron reactions might be of interest for 
experiments having to shield $\gamma{}$-rays as well as neutrons. 

\par
The same simulations can also be used to quantify single 
material performance for pure neutron shielding. For this
purpose, it is most revealing to study the energy-dependence of the neutron 
flux as it progresses through a given bulk material as a function of depth. This 
way, depletion of neutrons by absorption and 'feeding' of low energy 
neutrons by moderation of higher energy ones can be demonstrated. 
\par

\begin{figure}
\begin{center}
\includegraphics[width=14cm]{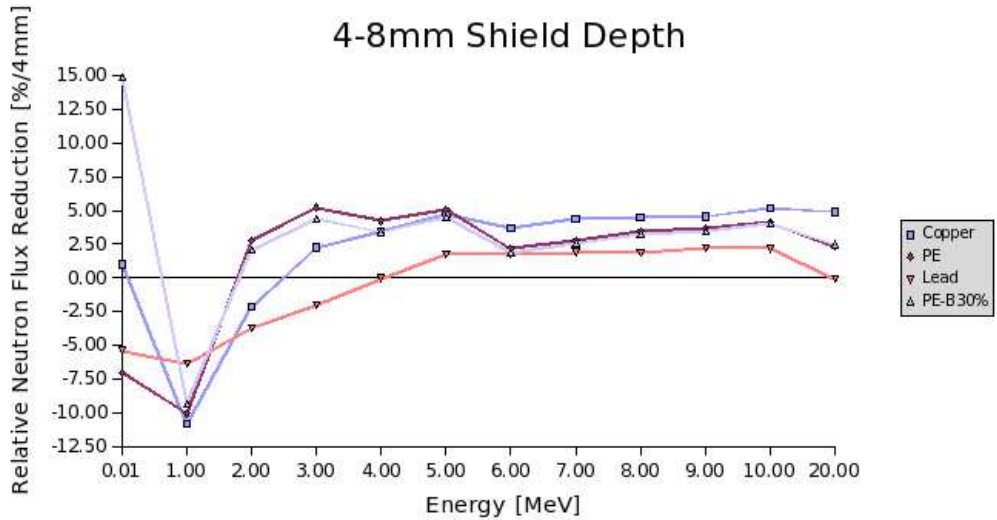}
\end{center}
\caption{Relative flux reduction per energy bin for copper,
  PE, lead and PE-B30\% as
bulk shielding materials. A Gran Sasso neutron flux input spectrum is made to impinge on a 10cm total thickness slab of material in the form of
a point-like directed beam source. The result at a depth of 4-8mm material
thickness is shown. Lateral dimensions are large
compared to the thickness. The
neutron flux is tallied at the surface. The relative
flux reductions result from flux changes through the
surfaces. A reduction of
flux for a given energy bin results in a positive value and flux
enhancement in a negative value. The flux changes are all normalised to
the minimum slab thickness of 4mm. The energy bin values contain 
flux up to the energy value, i.e. the first bin integrates flux from
thermal energies up to 10 keV, then from 10keV up to 1 MeV etc.\label{fig3}}
\end{figure}

\begin{figure}
\begin{center}
\includegraphics[width=14cm]{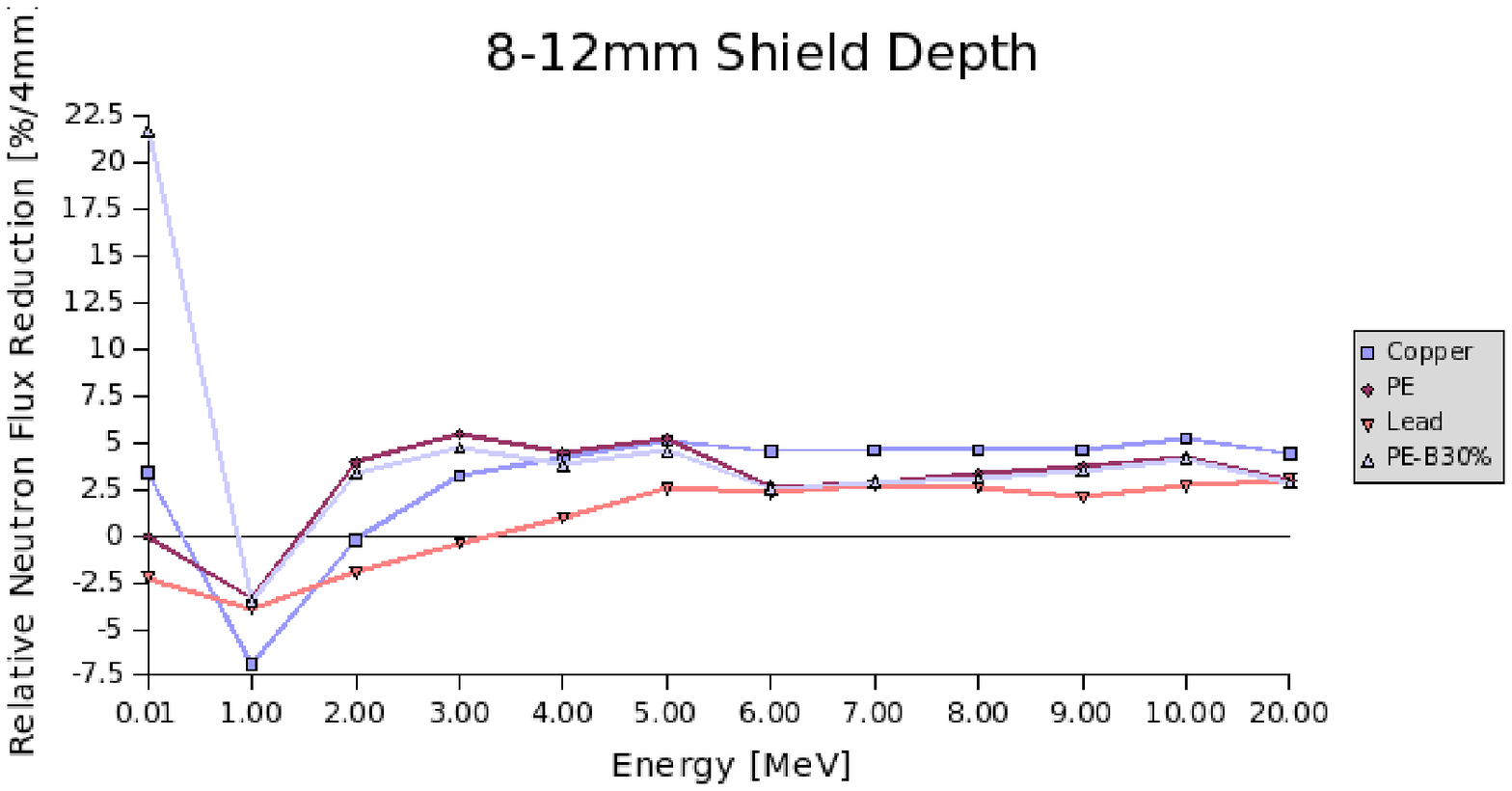}
\end{center}
\caption{As in figure \ref{fig3} except the result shown is for a depth of
  8-12mm. \label{fig4}}
\end{figure}

\begin{figure}
\begin{center}
\includegraphics[width=13cm]{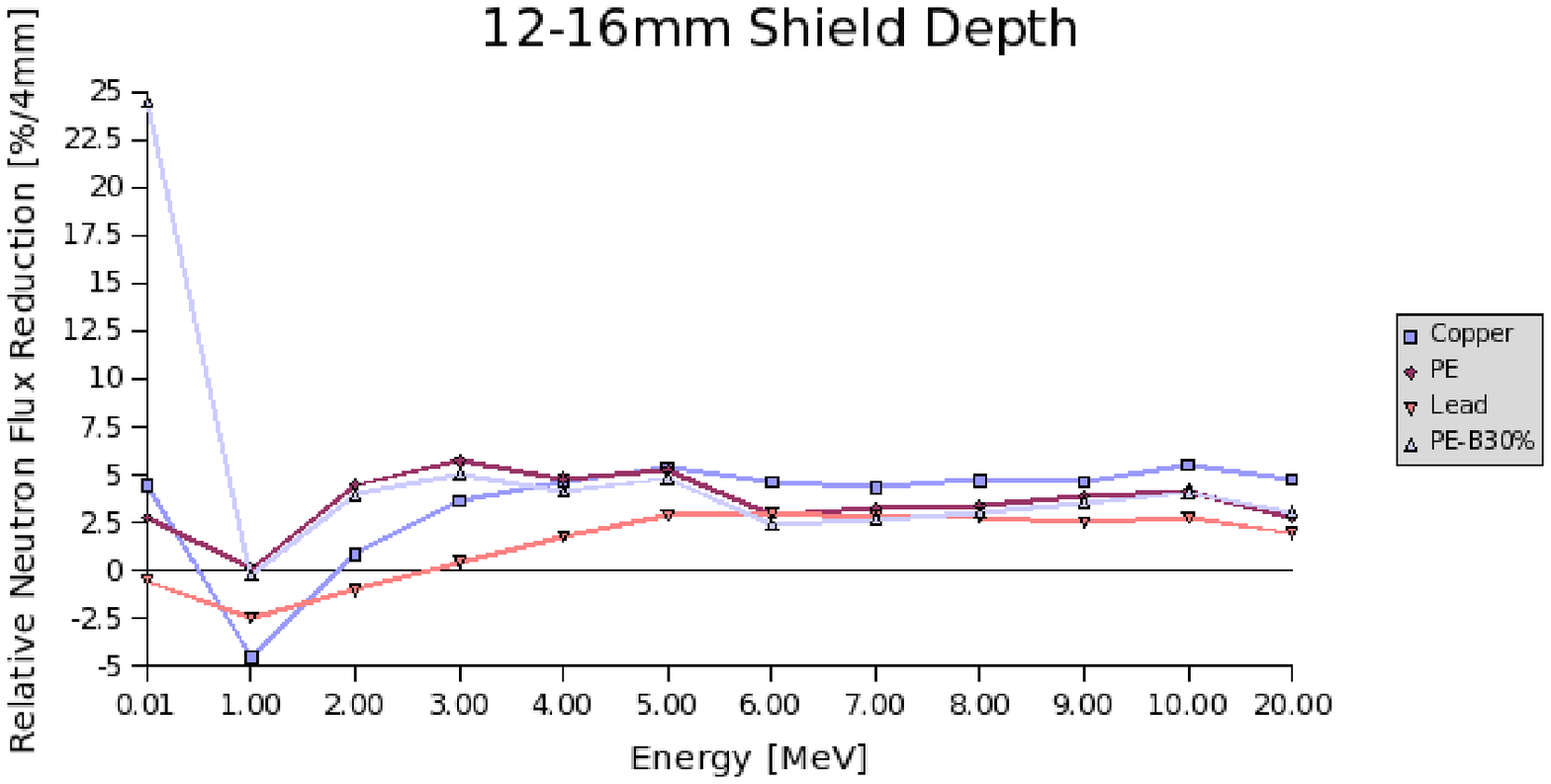}
\end{center}
\caption{As in figure \ref{fig3} except the result shown is for a depth of
  12-16mm\label{fig5}}
\end{figure}

\begin{figure}
\begin{center}
\includegraphics[width=13cm]{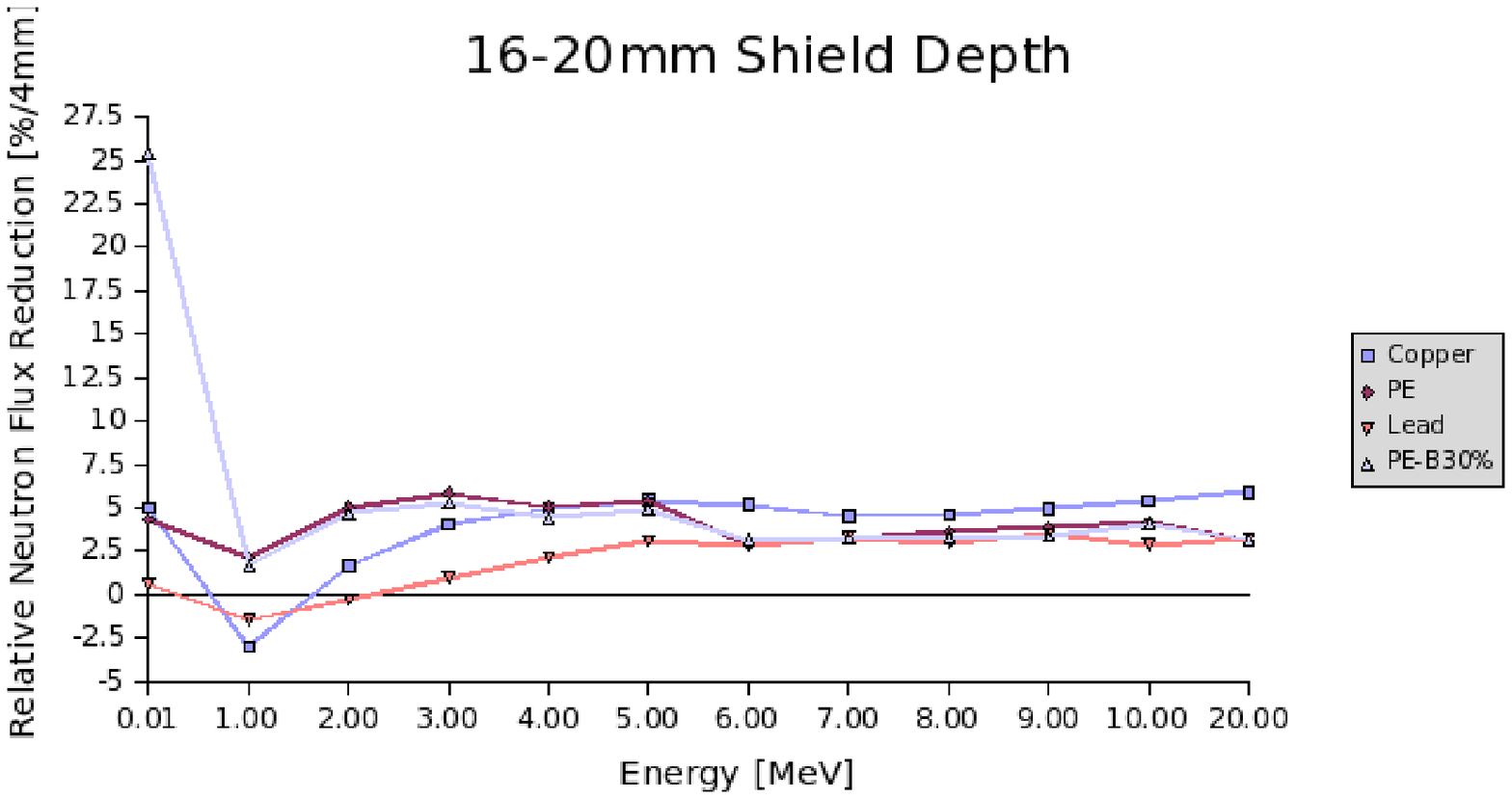}
\end{center}
\caption{As in figure \ref{fig3} except the result shown is for a depth of
  16-20mm\label{fig6}}
\end{figure}

\begin{figure}
\begin{center}
\includegraphics[width=13cm]{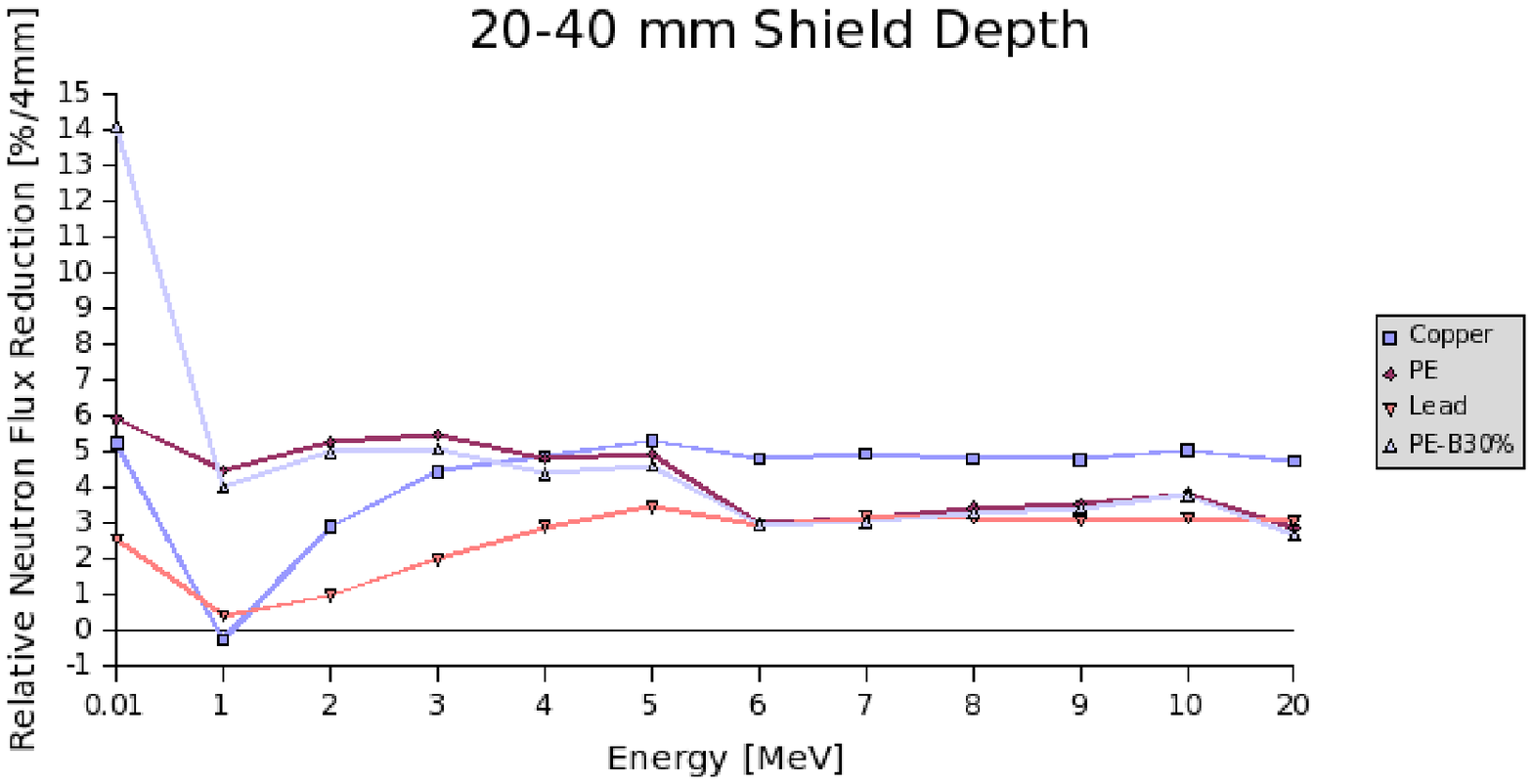}
\end{center}
\caption{As in figure \ref{fig3} except the result shown is for a depth of
  20-40mm\label{fig7}}
\end{figure}

\begin{figure}
\begin{center}
\includegraphics[width=13cm]{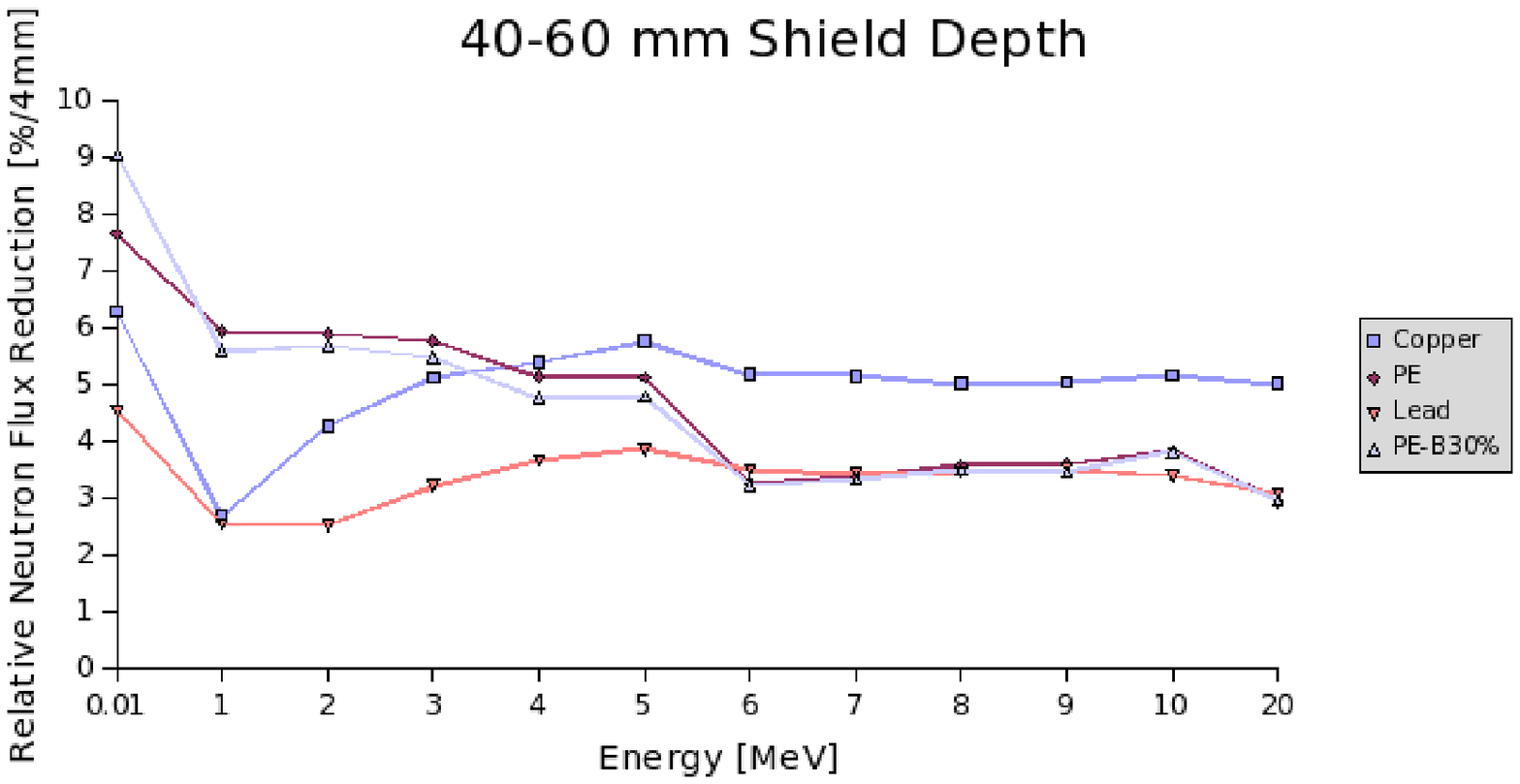}
\end{center}
\caption{As in figure \ref{fig3} except the result shown is for a depth of
  40-60mm\label{fig8}}
\end{figure}

\begin{figure}
\begin{center}
\includegraphics[width=13cm]{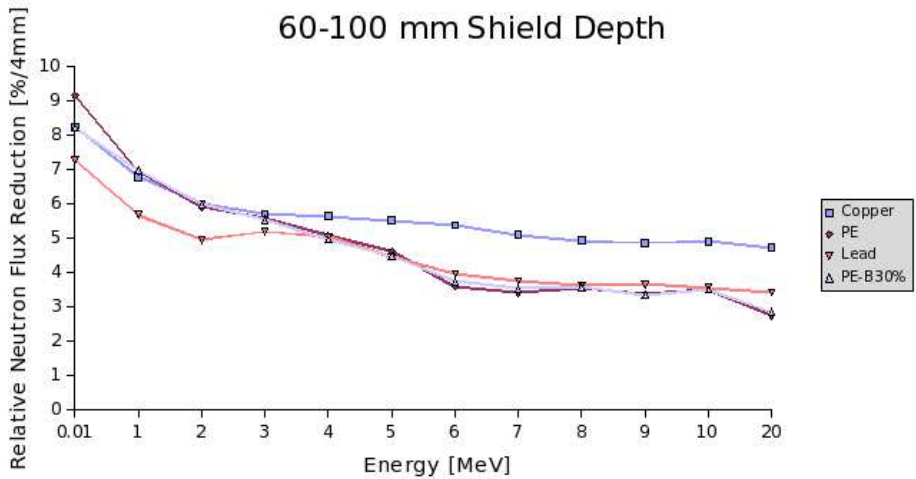}
\end{center}
\caption{As in figure \ref{fig3} except the result shown is for a depth of
  60-100mm\label{fig9}}
\end{figure}

\par
Figures \ref{fig3} to \ref{fig9} show the performance of the most
relevant  
materials in radiation shields including pure
polyethylene  
and a special neutron capture and moderator combination material, 
PE-boron (30\% B content). Each figure shows the reduction in neutron
number flux as a function  
of energy for bulk material slices of 4mm. As before, a neutron beam with Gran 
Sasso energy spectrum is transported through a 10cm thick wall of
material. 
The charts show a positive y-value for the relative flux change between
indicated surfaces 
(i.e. the case where neutron number is reduced) for the specified energy interval,
and a negative  
value in the case where enhancement of neutron number takes place. The neutron flux is
tracked for  
each 4mm until a depth of 20mm has been reached, then in steps of 20mm 
up to a depth of 60mm and finally, the last step to the exit--face
of the wall  
amounts to 40mm. 

\par
Something to bear in mind when studying figures \ref{fig3} to
\ref{fig9} is that the reduction in each 4mm slice is as a percentage
of the number of neutrons incident on that particular slice, rather
than of the original number incident on the outermost layer. The figures must therefore be
studied with care so that a reduction within a particular
slice is not confused with an overall reduction in the original number
incident on the outer surface of the slab of material. An example of where
this may happen is in the 1MeV energy bin for copper or lead. Initially there is an
overall increase in the number of neutrons in the outermost slices but eventually, at greater
depths in to the material, neutron numbers are reduced. The neutrons must
pass further through the slices before there are fewer neutrons in that energy bin than were intitially
incident on the outermost slab.

\par
Each of the materials presented display some unique features which can give 
important constraints on their usefulness as part of a neutron
shield. Lead  generally shows a relatively small influence on the neutron flux
which can
 be seen 
from the values of reduction or enhancement, respectively, compared with all 
other materials presented. In addition, a comparatively large
thickness of lead  
is necessary before the metal starts to reduce neutron flux over all energies. 
Sheets of lead with thickness below 2cm will rather
enhance neutron  
numbers, even at relatively high neutron energies below 2 MeV. The more 
common lead--brick with minimum thickness of about 5cm, however already 
reduces neutron fluxes significantly and provides a substantial self-shielding 
thickness against $(n,\gamma{})$-radiation (see Fig.~\ref{fig2}). Thick lead 
layers would even start to reduce efficiently a thermal neutron population.

\par
Copper (as well as Iron, not shown here) as a neutron shielding material
displays a remarkable thermal neutron  
capture cross section such that thin layers efficiently reduce a 
thermal neutron population. Nevertheless, a layer below 8mm thickness
would enhance the neutron 
flux between 1 MeV and 2 MeV. The overall reduction is significantly stronger 
than lead for higher energies (5-6\% compared to roughly 3\%). 

\par
The classic neutron moderator, polyethylene, changes the neutron
flux similarly  
to copper apart from the drastically different behaviour for
low-energy neutrons. Another, more subtle difference can be inspected
at the highest incoming energies, where 
the high density of copper results in stronger moderation of neutrons compared 
to PE. Below about 5 MeV, PE starts to scatter neutrons more effectively. 
Additionally for thicker slices, beyond 4cm, PE also absorbs such
moderated neutrons  
more and more, favouring thick shield sizes for PE as a
neutron shield material.  
Thin slices would inevitably lead to significantly enhanced neutron
populations at low energies. 

\par
A 30\% enrichment of PE with natural boron as thermal neutron absorber
and corresponding density increase, yields a significantly different
picture compared to pure PE. A strong reduction for low energy
neutrons can be seen for thin shield walls. At higher energies
the shape of the reduction as function of energy is practically
identical to pure PE.  Already thin slices of such enriched PE can
significantly change the composition of a neutron population. 

\par
Now taking into account these pieces of information, one might start
to compose  
a shield that most effectively attenuates neutrons and $\gamma{}$-rays
under all  
circumstances. The minimal thicknesses of shielding materials taken from this section can
be seen in Tab.~\ref{tab2}. 

\par
\begin{table}
\caption{Suggested minimal thicknesses of bulk material. The last
  column indicates the reason. Neutron attenuation means 
  thickness according to
  Figs.~\protect{\ref{fig3} to \ref{fig9}},
  whereas $(n,\gamma{})$ self-shielding comes from
  Fig.~\protect{\ref{fig2}} and is defined as that thickness where
  photon production probability reaches initial (after 4mm) values
  again, i.e. after the maximum.\label{tab2}}
\vspace*{6pt}
\begin{center}

\begin{tabular}{lll}
\hline
Material & Minimal thickness & criterium\\\hline
Lead & 20-40 mm& effective neutron attenuation\\
Lead & 70 mm& $(n,\gamma{})$ self-shielding effective\\
Iron & 20-40 mm& effective neutron attenuation\\
Iron & 14 mm& $(n,\gamma{})$ self-shielding effective\\
Copper & 40-60 mm& effective neutron attenuation\\
Copper & 16 mm& $(n,\gamma{})$ self-shielding effective\\
PE & 16-20 mm& effective neutron attenuation\\
PE & 30 mm& $(n,\gamma{})$ self-shielding effective\\
PE-Bi and PE-B& 12-16 mm& effective neutron attenuation\\
PE-Li & 8-12 mm& effective neutron attenuation\\
\end{tabular}
\end{center}
\end{table}
\par

\section{Multilayer Shield Configurations}\label{sec2}

Any neutron shield must be constructed such that higher energy
neutrons are moderated as fast as possible and subsequently captured
in order to significantly remove neutrons. An added complication for a
shield design can be the accompanying $\gamma{}$-radiation from capture
processes. Furthermore, a shield should be as compact and
cost-effective as possible. Combining these constraints was the main
purpose of this study. 

\par
\begin{figure}
\begin{center}
\includegraphics[width=9.0cm]{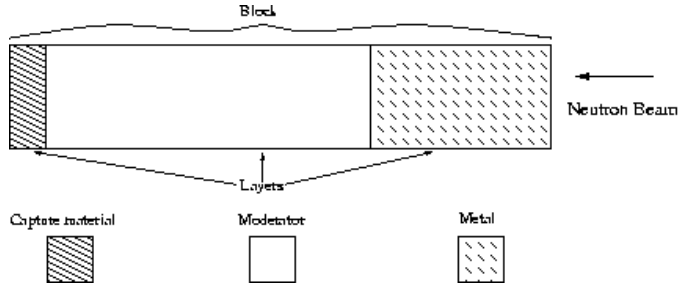}
\end{center}
\caption{Diagram of a single block made up of layers of a moderator,
  capture material and a metal. \label{fig12}}
\end{figure}

One idea that was investigated, having established a suitable combination and
configuration of materials, was testing small repeated blocks of layers, see Fig.\ref{fig12}, to see if it offers
a greater reduction in total neutrons and photons compared to a single large block of equal total thickness. The reasoning behind this was that
the blocks of layers may yield approximately exponential attenuation
of incoming flux at each block. In principle, this idea cannot be optimal
since after the first block, the flux has changed in shape
considerably such that a repetition might not be the best way to
remove incoming flux at that position. It was agreed that a multilayer configuration was worth
testing nevertheless. 

\par
Starting with a minimum layer depth for all realistically available
materials, see Tab. \ref{tab2}, a minimal block unit of layers was constructed after all
possible permutations (ordered with respect to neutron beam arrival
direction) of up to four materials had been put to test.

\par
The chosen method for the comparison
 of various configurations was to first fix the single block depth to
 15cm, leaving enough space to
accommodate thicknesses of single materials in the range of minimal
requirements. Second, a layer
configuration is arranged such that moderation and capture should
happen in series, see below and \cite{hong}. Capture materials in front of moderators have been
excluded from this study after preliminary simulations showed 
significantly inferior performance as expected. The remaining possible
variations comprise the placement of high-density materials for either
photon capture or neutron capture and moderation or even both. Three 
sets of configurations have been tested, either a metal behind or in
front of a moderator$+$capture structure or both. Best results for neutron attenuation were obtained for a  metal in front of the moderator$+$capture
material, see Fig.\ref{fig12}.

\par
An 80cm total shield thickness allowed 3 blocks of these layers to be sandwiched between 15cm lead and 20cm pure PE for simulations, see discussion on this 'clamp' below. The 3-block arangement was chosen as arbitrary. In all of the following tests, 10$^{8}$ particles with a Gran Sasso spectrum were simulated. 

\par
\begin{figure}
\begin{center}
\includegraphics[height=10cm,width=20.0cm, angle=90]{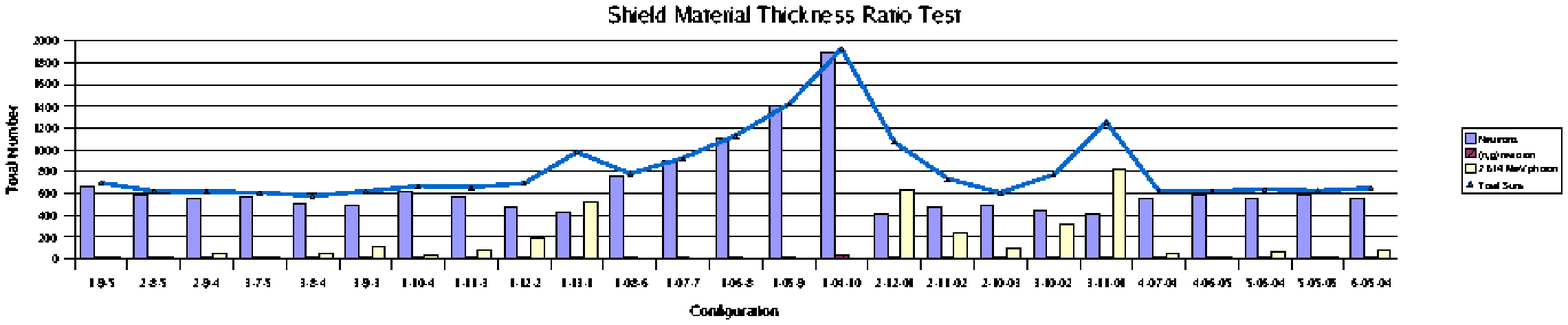}
\end{center}
\caption[angle=90]{25 different ratio configurations of material thickness were tested using Lead as the metal, PE-Bi as the moderator and PE-Li as the capture material. The best configuration for overall particle reduction is shown to be 3:8:4 (3cm PE-Li, 8cm PE-Bi, 4cm Lead). \label{figratiopics}}
\end{figure}

\par
Using lead as the metal, PE-Bi as the moderator and PE-Li as the capture material as an initial best guess, tests were performed to optimise the ratio of one material thickness to another. In total, 25 ratio configurations were tested and a 3:8:4 configuration (3cm PE-Li, 8cm PE-Bi, 4cm Lead) was seen to be best for an overall reduction of both neutrons and gammas, although this was not best for neutron or gamma reduction alone, see Fig.\ref{figratiopics}. A 2:12:1 configuration was best for neutron reduction and any configuration with 9cm of lead or more removed all gammas (excluding those from (n,$\gamma$) reactions). 

\par
The reader should bear in mind that the moderator material will follow the pattern presented in Fig.\ref{fig1} for neutron reduction versus thickness but the capture material will behave differently due to the fact the moderator has had an effect prior to reaching the capture material surface. In general, a 1cm increase in PE-Li at the expense of 1cm in PE-Bi will have a greater effect on number reduction due to the small amount of PE-Li simulated in comparison to PE-Bi. A small amount of PE-Li is tested as it is not sold in large blocks, although it is generally cheaper than PE-Bi. It appears that 4-5cm of lead is optimal at these scales as further thickness increase has little more effect on gamma number and (n,$\gamma$) reactions have a minimal contribution to total numbers so may be ignored.

\par
Following the material ratio testing, further simulations were run in order to find the best combination of materials. The metal was exchanged between lead and iron for comparison whilst the moderator was swapped between PE and Bi-PE. Results show that Bi-PE is the better moderator and that lead is better in overall particle reduction. PE-Li was compared to PE-B and was found to out-perform as a capture material. As a single material, lead is more effective than iron at gamma attenuation but iron is more suited to neutron reduction. 

\begin{figure}
\includegraphics[width=14.0cm]{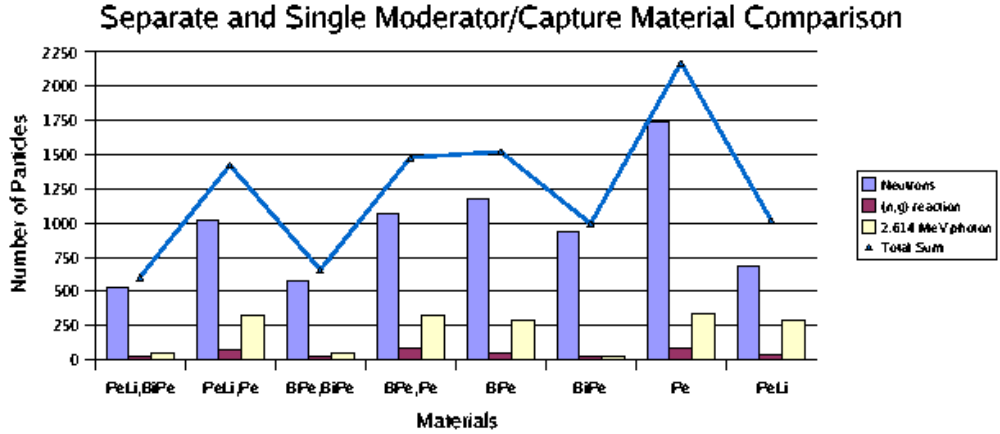}
\caption{Comparison between single capture/moderator materials and a
  separate moderator then capture structure. Each structure follows
  the 3:8:4 pattern, as previously used, with lead as the metal in
  each case. In the case where a single material is used, the total thickness of moderator and capture material is taken i.e. 11cm.\label{modcapmaterial}}
\end{figure}

\par
Another question that was addressed was whether it is better to have a separate moderator-capture configuration or to combine the two tasks using a single material. The results of simulations show a definite argument towards a separate moderator and capture material for overall particle reduction, see Fig. \ref{modcapmaterial}. The best material combination was using PE-Bi for the moderator and PE-Li for the capture material.

\par
\begin{figure}[t]
\includegraphics[width=14.0cm]{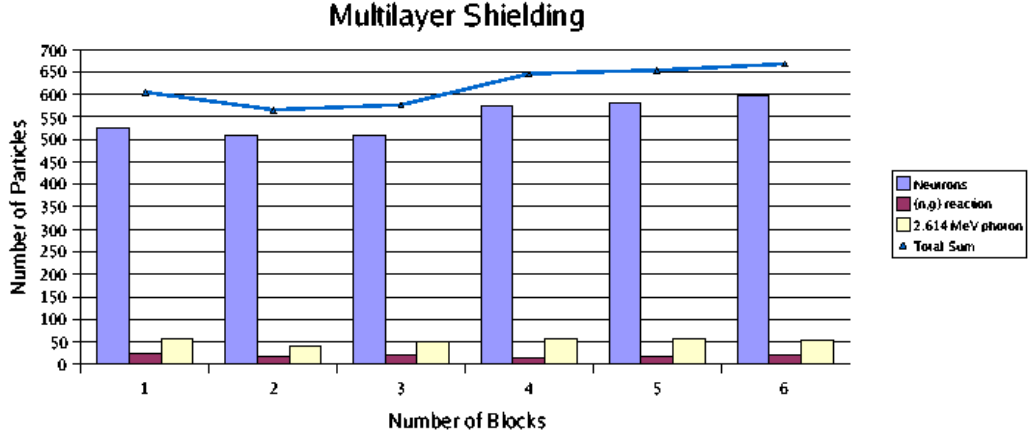}
\caption{Comparison of results for varying the size and therefore the number of blocks sandwiched between 15cm lead and 20 cm PE giving a total shield thickness of 80cm. The ratio of materials in each block is kept at 3:8:4 (PE-Li:PE-Bi:lead).\label{figmultilayer}}
\end{figure}

The next stage in the testing was to then find out whether a single
large 
layered block would achieve better or worse results than repeated
smaller blocks of the
3:8:4 configuration, both sandwiched between 15cm lead (inside) and
20cm pure PE (outside). This was done by fixing the total shield
thickness to 80cm, keeping the 3:8:4 ratio constant and adjusting the
number of blocks in the sandwich by varying the block size
proportionally. The outcome of this can be seen in Fig. \ref{figmultilayer}.

\par
Results show that a repetition of two blocks reduced the total number of particles passing through the shield, although not significantly, see Fig \ref{figmultilayer}. There was very little difference between a single elongated block and  two smaller blocks with the same material thickness ratio; the results for the two blocks were within percentage errors. Further repetitions of blocks had a detrimental effect. 

\par
A subsequent test to see whether an outermost pure moderator layer 
would enhance the multilayer shield, showed that this improves overall performance considerably due to the initial
softening of any incoming neutron spectrum. Both default layers (introduced as educated guess structures initially),
bracketing the multilayer structure, do not need to be
thick. An additional 5cm inner lead (should be low in $^{210}$Pb as
usual, see \cite{heusser}) would already remove almost the entire
photon flux from any neutron interaction. The softening moderator
outside can be a 10-20cm pure PE layer, representing a compromise in
moderation efficiency and in cost and weight. 

\par
Experiments that rely on removing
environmental photon flux entirely should always aim at a
total of 30cm lead for the 2.614 MeV line from the Thorium
chain. Therefore it might be necessary to increase the inner lead
layer to a total of 15cm. 

\par
Considering that this proposed multilayer structure is clamped between
a very conventional shield configuration (PE outside, lead inside) the
question arises whether it is necessary at all. This topic is
discussed in the next section.

\section{Comparison with Standard Shields}\label{sec3}

\begin{figure}
\includegraphics[width=14cm]{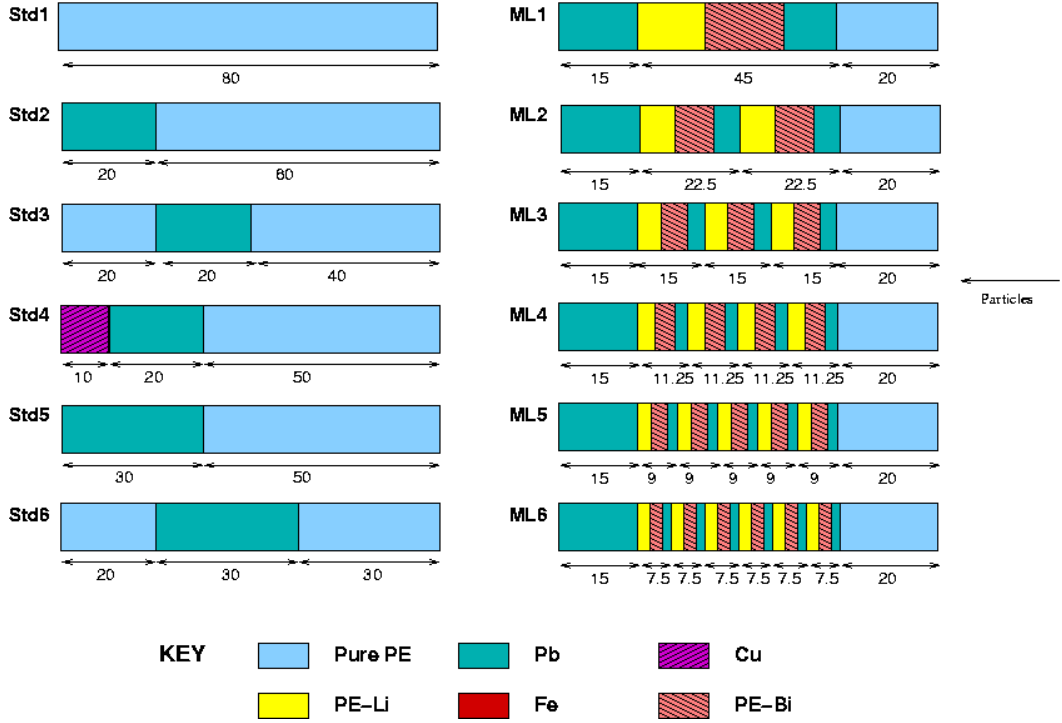}
\caption{Shielding configurations simulated for comparison of standard
shields (label: Std) to proposed multilayer shields (label: ML). Each of the ML shields keep the thickness ratio of PE-Li:PE-Bi:Pb to 3:8:4 as this was shown to be the best configuration from the simulations. For
discussion, see text.\label{fig10}}
\end{figure}

In order to gauge how well the shields in the previous section (ML1-6) would perform compared to a selection of six existing shielding designs (Std1-Std6), simulations using MCNPX were run, see Fig.~\ref{fig10}. A Gran Sasso flux beam consisting of 10$^{8}$ neutrons was started as well as a mono-energetic (2614.5 keV) gamma-ray beam source. The performance of each shield was based on how many neutrons and gammas would leave the exit face of the shielding. For all
configurations a total thickness of 80cm has been set, determined by
the proposed configurations.

\par

\par

The pure PE shielding (Std1) is considered in \cite{drift1,drift2} as a
viable shielding option although with a thinner total depth. The main difference between Std2,3 and Std5,6 is merely the lead thickness. Std2 and
Std5, respectively, have the classic PE-lead configuration, where it
should be pointed out that 30cm lead is not unusual but 50cm PE is
considered to be a very massive moderator \cite{heusser}. The
configurations Std3 and
Std6 are more recent actual shields employed in dark matter
experiments but with different, thinner layer depths. Here the
inner moderator is considered to be an effective method to moderate
neutrons originating from the shielding material itself, in particular
from the
lead layer by neutron spallation reactions. The Std4 shield is
another conventional shield configuration, seen in dark matter as well
as double-beta searches. The copper layer in thinner form would
typically represent an inner lining to remove lead x-rays
\cite{heusser}.  

\par
\begin{figure}
\includegraphics[width=14.0cm]{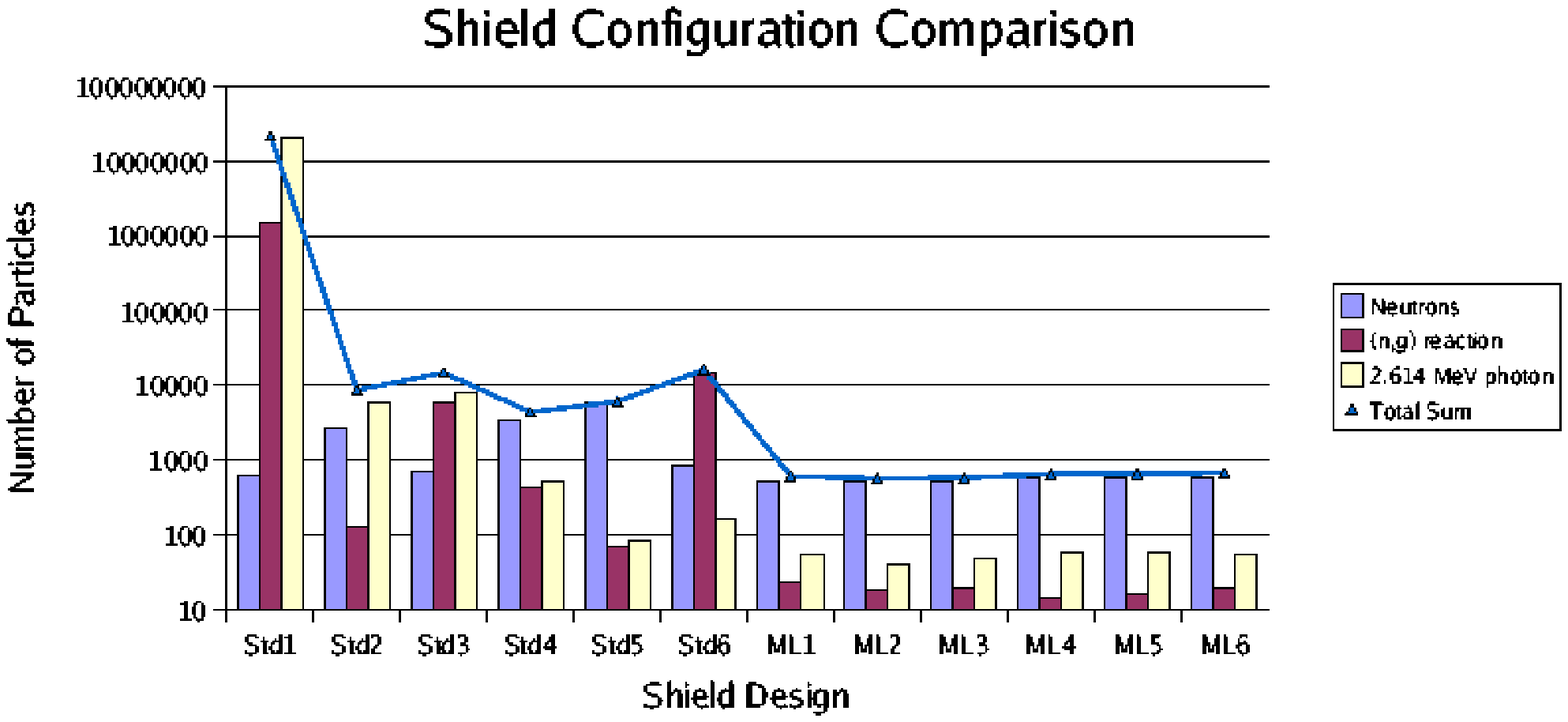}
\caption{Shield configuration comparison for six standard shields,
  labeled 'Std' and six multilayer shields, labeled 'ML'. The
  configurations can be seen in Fig.~\protect{\ref{fig10}}. Shown here is
  the total number of either neutrons or photons tallied at the exit
  surface of the respective shield configurations. The first bar shows
neutron numbers and the second photon numbers from
$(n,\gamma{})$-reactions. Those result from a Gran Sasso neutron flux
beam source, starting $10^{8}$ neutrons. The third shows photon
numbers from a monoenergetic (2614.5 keV) $\gamma{}$-ray beam source,
starting $10^{8}$ photons. The line indicates values for the total sum
of neutrons and photons traversing the shield.\label{fig11}}
\end{figure}

\par
As discussed in the previous section, the proposed shields contain 
a number of consecutive blocks of multilayers, clamped between a conventional Lead-PE structure (ML1-6 in Fig.\ref{fig10}). The results of the simulations can be seen in Fig.\ref{fig11}. It can be seen that even the poorest of performers from ML1-6 still exceeds performances from all of the existing shields that were tested. 

\par
The standard shields exhibit some interesting features. Note that
the pure neutron fluxes of Std1,3 and 6 are comparable although
moderator thicknesses vary from 80 to 50cm total. Disturbing the
effect of pure PE on neutrons by implementation of lead has a
favourable consequence on neutron attenuation (compare Std2,3 and
Std5,6). However, the effect on the photon population is
opposite. The build-up factor and $(n,\gamma{})$-reactions increase the
total considerably and strongly favour lead as last layer. Replacing
the last 10cm of lead with copper (Std5 to 4) also shows a
non-negligible effect on the photon population. 

\par
Finally, it might be interesting to note that the multilayer shields exhibit
similarly strong attenuation 
of neutrons to the pure PE shield. The added multilayer advantage of more
than four orders of magnitude suppression of photon flux compared with the
pure PE shield might be of interest, even for quasi $\gamma{}$-ray
insensitive detectors. 
\par

\section{Conclusions}\label{end}

A survey of available radiation shielding materials has been
presented. An effective configuration for combined neutron,
photon attenuation was derived and compared to existing configurations for
radiation shields in underground nuclear and particle physics
experiments. For a total background reduction in this kind of
experiment it is proposed that multilayer shields composed of a metal, moderator and capture material, as in Fig.\ref{fig12}, are used.  The multilayer
structure turns out to be more efficient in dealing with neutron flux
evolution. Thermal neutron absorbers have been shown to be very
efficient already when used as thin sheets rather than as
significantly more expensive bulk material. For pure moderation
purposes, simple PE is 
sufficient. Metals are necessarily present for $\gamma{}$-ray
attenuation but have been shown to be interesting also for their
influence on neutron fluxes. 

\par
Finally, it should be pointed out that this study was concerned with
passive shielding only. The neglected background contributions due to
muons, even in underground laboratories, are best dealt with using active
shielding components such as plastic scintillator panels mounted outside
the shielding (see e.g. \cite{heusser,schroettner}). For
COBRA, improvements on the passive shielding design have resulted in the development of an active element which could be mounted as an inner layer. An active element in this position can act as a veto for any last remaining external
background. Further
studies examining the minimal passive shield configurations for COBRA, subject to cost constraints,
are ongoing.

\end{document}